\documentclass[referee]{raa}            
\usepackage[none]{hyphenat}

\usepackage{graphicx,times}             
\usepackage{float}

\usepackage{natbib}

\usepackage{xfrac} 
\usepackage{siunitx}
\usepackage[pagebackref=true]{hyperref}
\usepackage{amssymb,amsmath}
\bibpunct{(}{)}{;}{a}{}{,}
\usepackage{siunitx}

\usepackage{overpic} 

\begin{document}

  \title{Suzaku Observation of Merging Clusters Abell 222 and Abell 223}

   \volnopage{Vol.0 (20xx) No.0, 000--000}      
   \setcounter{page}{1}          

   \author{Yanling Chen  
      \inst{1}
   \and Wei Cui
      \inst{1}
   \and Aurora Simionescu
      \inst{2}
   \and Rui Huang
      \inst{1,3}
   \and Dan Hu
      \inst{4}
   }

   \institute{Department of Astronomy, Tsinghua University, Beijing 100084, China; {\it yl-chen19@mails.tsinghua.edu.cn}\\
        \and
             SRON Netherlands Institute for Space Research, Niels Bohrweg 4, 2333 CA Leiden, The Netherlands\\
        \and
             Department of Astronomy, University of Michigan, Ann Arbor, MI, 48109-1107, USA\\
        \and
             Department of Theoretical Physics and Astrophysics, Faculty of Science, Masaryk University, Kotlářská 2, Brno, 611 37, Czech Republic\\
\vs\no
   {\small Received 20xx month day; accepted 20xx month day}}
   
\abstract{Previous X-ray and optical studies of the galaxy cluster pair Abell 222/223 suggested the possible presence of a filamentary structure connecting the two clusters, a result that appears to be supported by subsequent weak-lensing analyses. This filament has been reported to host a primordial warm-hot intergalactic medium (WHIM), which existed prior to being heated by the interactions of the clusters. In this study, we made an attempt to examine the reported emission feature with data from an archival Suzaku observation, taking advantage of its low detector background. Because the emission is expected to be very weak, we first carefully examined all potential sources of ``contamination", and then modelled the residual emission. Due to large uncertainties, unfortunately, our results can neither confirm the presence of the reported emission feature nor rule it out. We discuss the sources of uncertainties.
}

   \authorrunning{Y.-L Chen et al.}            
   \titlerunning{Suzaku Observation of Abell 222/223}  

   \maketitle

%
%
\section{Introduction}           
\label{sect:intro}

The cosmic missing baryon problem refers to the discrepancy between the observed baryon content in the local universe and the measurements made with the observations of the Cosmic Microwave Background, primordial elemental abundances and the Lyman alpha forest (\citealt{CenOstriker99}). Cosmological simulations (e.g., \citealt{CenOstriker99, Tuominen21}) show that the ``missing'' baryons are predominantly located in large-scale filamentary structures. Such diffuse gas is often referred to as warm-hot intergalactic medium (WHIM), because it is of temperature ranging from $10^5$ to $10^7$\,K (\citealt{Dave01}). Much progress has been made on detecting the warm component of WHIM with O~VI absorption line observations (\citealt{Shull12}). However, the hot component of WHIM is still poorly measured.

The detection of the hot component of WHIM is highly challenging in the soft X-ray band (\citealt{Cen06}). Nevertheless, extensive effort has been made to carry out absorption-line observations of WHIM with X-ray gratings onboard \textsl{Chandra} and \textsl{XMM-Newton}, and positive detections have been reported (see \citealt{Fang24} for a review), albeit limited to only a few lines of sight. Attempts have also been made to detect X-ray emission from hot WHIM in a filament that connects a pair of clusters, with some success (\citealt{Sato10, Mirahor22, Alvarez18, Reiprich21, werner08}). The A222/223 system is of particular interest in this regard. The X-ray emission from the filament connecting the two clusters was reported \citep{werner08}, based on \textsl{XMM-Newton} observations of the system, and the detection significance was said to reach about 5$\sigma$. The presence of the filament was supported by a weak-lensing study, with the significance of the detection also exceeding 5$\sigma$ (\citealt{dietrich12}). Neither of the two member clusters is seemed to be relaxed, suggesting that interaction such as merging or sloshing is taking place in the system (\citealt{durret10}).

In this work, we analyzed archival data from the \textsl{Suzaku} observations of A222/223, making use of its low particle background to further examine the filament emission. The structure of this paper is as follows: In Sect.\ref{sect:data} we describe the \textsl{Suzaku} observations and data analysis. In Sect.\ref{sect:results}, we present constraints on the properties of any WHIM emission from the \textsl{Suzaku} data analysis. In Sect.~\ref{sect:discussion}, we discuss our findings and compare them with previous studies. Unless otherwise stated, all errors throughout this paper are quoted at the $1 \sigma$ confidence level. The cosmological model adopted in this work is the standard $\Lambda$CDM cosmology, with $H_0 = 70\ \mathrm{km\ s^{-1}\ Mpc^{-1}}$, $\Omega_M = 0.3$, and $\Omega_{\Lambda} = 0.7$. In this framework, the angular scale is 3.427$''$ per 1\,kpc, calculated at the redshift of 0.21, corresponding to the A222/223 system. The quantities $r_{200c}$ and $r_{500c}$ represent the radii in which the mean enclosed density is 200 and 500 times the critical density of the universe, respectively.

\section{Data analysis}
\label{sect:data}
\subsection{Suzaku Observations}
\label{sect:obs}

\begin{figure} 
   \centering
   \includegraphics[width=12.0cm, angle=0]{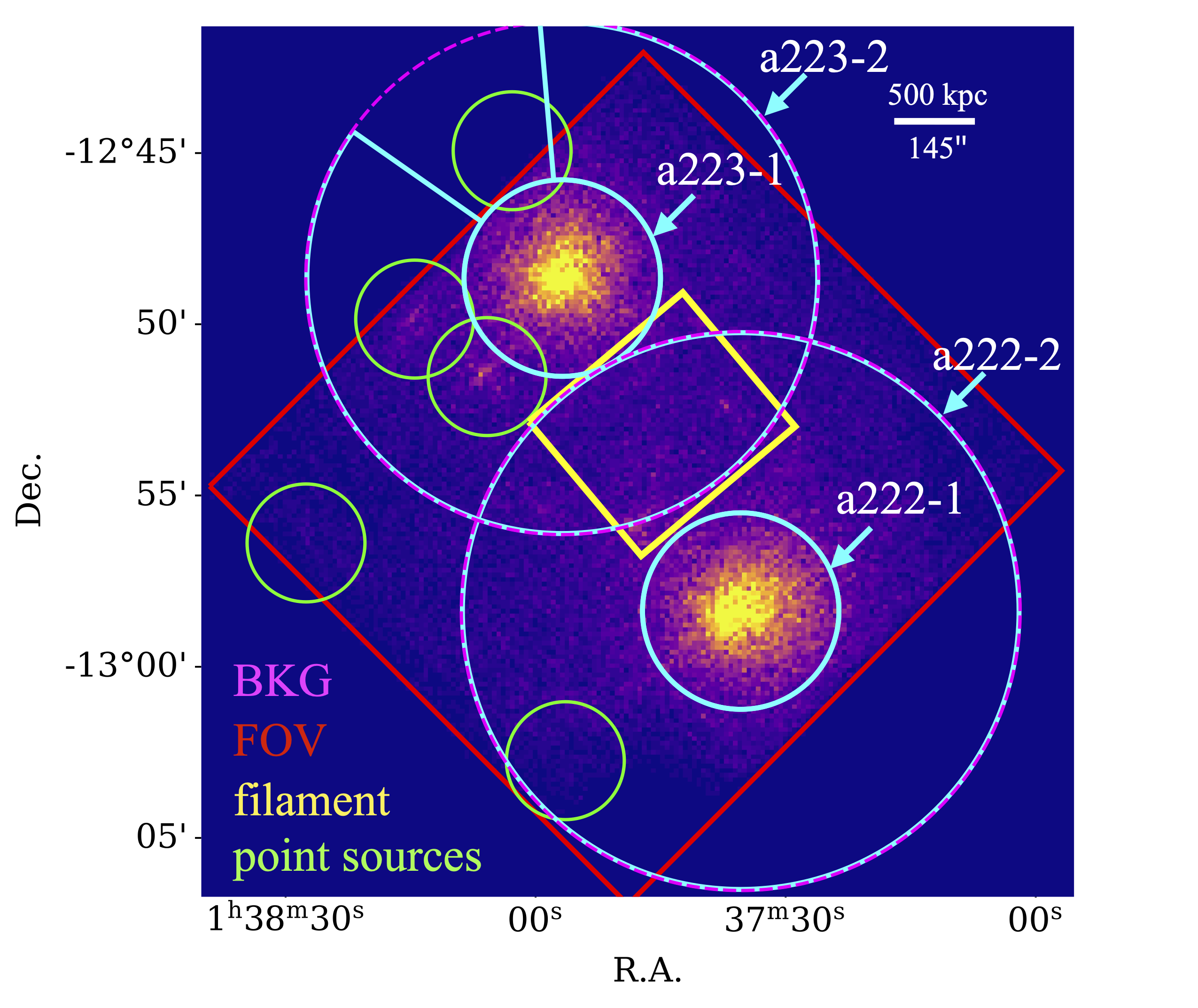}
   \caption{The 0.5-7.0\,keV vignetting-corrected \textsl{Suzaku} image of the A222/223 galaxy cluster pair. The cyan sectors and yellow box are the spectrum extraction regions for the clusters and the filament, respectively. The magenta dashed circles represent the $1.1r_{200c}$ of the two clusters, while the region outside the magenta circles is defined as the background region. The green circles mark the exclusion regions for point sources in the \textsl{Suzaku} data. The red box is the \textsl{Suzaku} FOV. The image was binned by 8 pixels, 
   and then smoothed using a 12.5$''$ Gaussian kernel.}
   \label{fig:regions}
   \end{figure}

Abell 222/223 was observed by the \textsl{Suzaku} telescope on December 25, 2010, as shown in Tab.\ref{tab:obs_information}, along with the \textsl{Chandra} and \textsl{XMM-Newton} observations. Fig.\ref{fig:regions} presents the \textsl{Suzaku} view of the pair of A222/223 clusters, where A223 is in the northeast and A222 is in the southwest, with their X-ray peaks at coordinates (RA, Dec) of (01:37:33.5, -12:59:20.9) and (01:37:55.9, -12:49:21.2), respectively. The virial masses $M_{200c}$ of the two clusters, as derived from weak lensing analysis, are determined to be $M_{200c}\left(\rm A222\right) = \left(2.7^{+0.8}_{-0.7}\right) \times 10^{14} M_{\odot}$ and $M_{200c}\left(\rm A223\right) = \left(3.4^{+1.3}_{-1.0}\right) \times 10^{14} M_{\odot}$ (\citealt{dietrich12}). To assess excess emission in the filament region, we carefully selected regions to quantify contribution from known sources, as shown in Tab.\ref{tab:reg_information}. We defined sector regions for cluster emission, centered on the X-ray surface brightness peaks, with a222-1 and a223-1 representing the core regions and a222-2 and a223-2 the outskirts (shown in cyan). Sector radii were chosen to 
be sufficiently large to account for scattered cluster emission. The outer boundaries of both cluster regions extend to 1.1$r_{\rm 200c}$,
where $r_{\rm 200c}$ are $1.28\pm 0.11$\,Mpc and $1.55 \pm 0.15$\,Mpc for A222 and A223, respectively. 
Here, we defined the filament region (fila-box) as a $6.1'\times5.3'$ rectangular box positioned on the merging axis connecting the two clusters (marked in yellow), fully encompassed by the outer boundaries of the cluster outskirt regions. This configuration allows an assessment of contribution of the cluster outskirt emission to the filament region. To avoid double accounting, we carefully separated the filament, a222-2, and a223-2 regions in their definitions. A sky-background region (bkg) was defined as the area outside the magenta circles marking the $1.1 r_{200c}$ boundaries of both clusters.
The point sources with 2-8 keV flux exceeding $1.8\times 10^{-14}~\rm erg\ s^{-1}cm^{-2}$ 
are marked in green circles and are removed from subsequent analyses. For \textsl{Suzaku} data pre-reduction, we generally followed the official pipeline of the X-ray Imaging Spectrometers (XIS)\footnote{https://darts.isas.jaxa.jp/astro/suzaku/analysis/process/v2changes/criteria-xis.html}. We used the standard screened event files from the observation, and combined events from both the $5 \times 5$ and $3\times 3$ clocking modes. The cutoff rigidity (COR), which was set to a threshold of 8\,GV instead of 6\,GV. After the screening, we examined the light curves of the observations and excluded the time intervals with a count rate exceeding 3$\sigma$, which returns 82\,ks effective exposure time for the XIS0/1/3 detectors. Since the vignetting effect was not accounted for by the \texttt{xisexpmapgen} task, we generated a vignetting map using \texttt{XISSIM}, simulating the vignetting effect by inputting a flat-field image. Subsequently, we multiplied the exposure map by the normalized vignetting map to generate the effective exposure time map\footnote{https://heasarc.gsfc.nasa.gov/docs/suzaku/analysis/expomap.html} and divided the counts map by it to generate the count rate map.

\begin{table}
\bc
\begin{minipage}[]{100mm}
\caption[]{
X-ray Observations of A222/223
\label{tab:obs_information}
}\end{minipage}
\setlength{\tabcolsep}{1pt}
\small
 \begin{tabular}{lcccc}
  \hline\noalign{\smallskip}
Telescope & Date & ObsID & Instruments & Exposure Time (ks)\\
  \hline\noalign{\smallskip}
\textsl{Suzaku}& 2010-12-25 &805035010&XIS0/1/3&101\\
\textsl{XMM-Newton}& 2007-06-18 & 0502020101 / 201&EPIC&144\\
\textsl{Chandra} & 	2004-09-10 / 2007-07-27 & 4967 / 7705&ACIS-I&50\\
  \noalign{\smallskip}\hline
\end{tabular}
\ec
\end{table}

\begin{table}
\bc
\begin{minipage}[]{100mm}
\caption{Spectral extraction regions \label{tab:reg_information}}\end{minipage}
\setlength{\tabcolsep}{1pt}
\small
 \begin{tabular}{lcccccc}
  \hline\noalign{\smallskip}
  & a222-1 & a222-2 & a223-1 & a223-2 & fila-box & bkg\\
  \hline\noalign{\smallskip}
Radius [arcmin]& 0 - 3 & 3 - 8.5 & 0 - 3 & 3 - 7.8 & 6.1 $\times$ 5.3 & \\
Area [arcmin$^2$] & 28.24 & 90.57 & 22.47 & 39.54 & 32.73 & 26.90\\
  \noalign{\smallskip}\hline
\end{tabular}
\ec
\end{table}

\subsection{Effects of scattered light}
\label{sect:clus-scat}

Because the primary objective of this work is to detect excess emission 
in the filament region, which is expected to be very faint (if any at all), we must carefully assess all sources of contamination. For Suzaku observations, scattered light is a major issue (\citealt{Ishisaki07}). 
We utilized data from \textsl{Chandra}, making use of its superior angular resolution, and the \texttt{XISSIM} simulator to compute scattered-light distributions. 

\begin{figure} 
   \centering
   \includegraphics[width=10cm, angle=0]{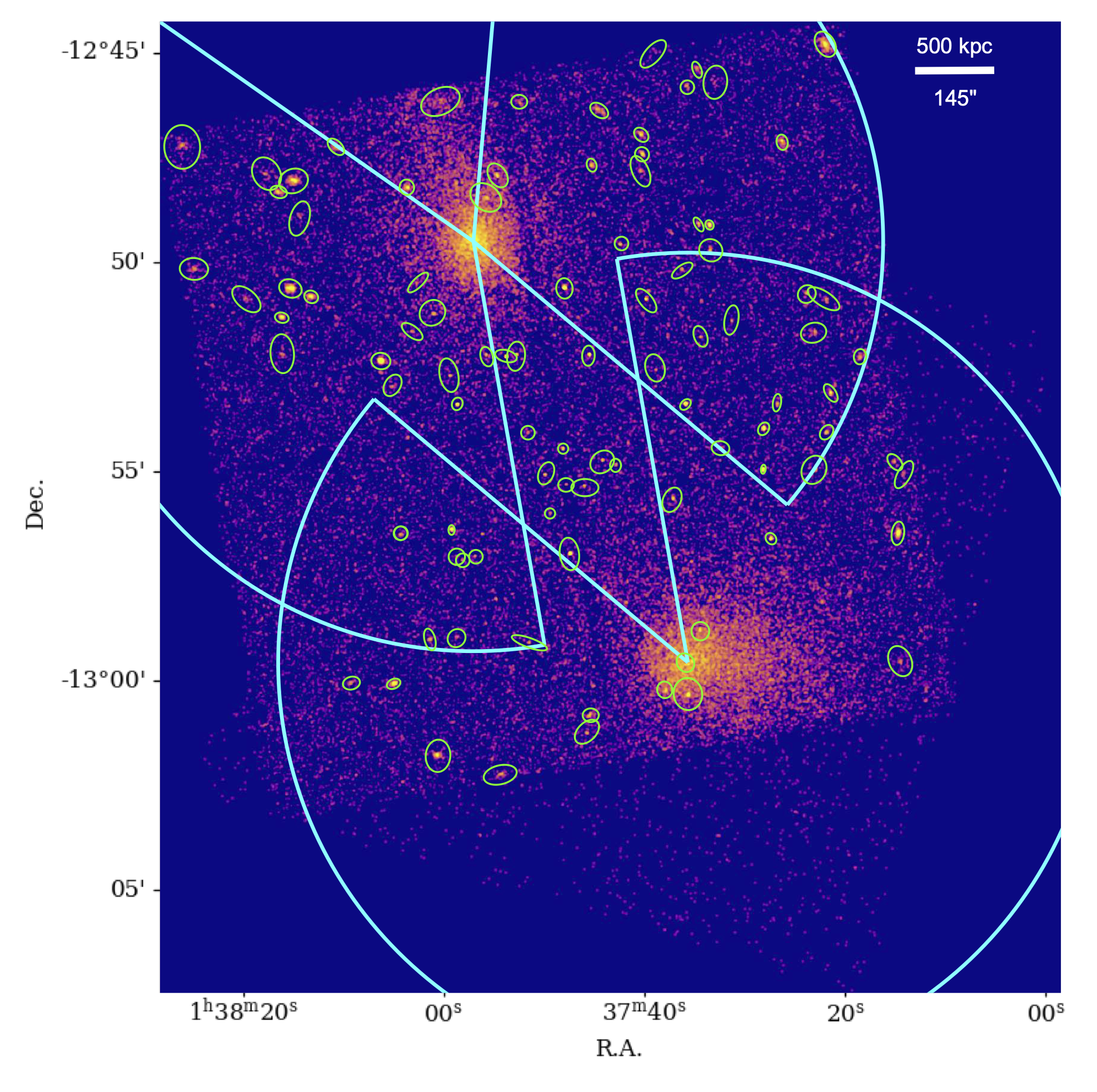}
   \caption{The \textsl{Chandra} image of A222/223 in the 0.5-7.0 keV band. The cyan sectors indicate the regions from which surface brightness profiles were extracted, while the green circles denote the exclusion zones for point sources. The image was smoothed using a Gaussian kernel with a width of 1.35$''$.} 
   \label{fig:clus_scat}
   \end{figure}

\begin{figure} 
   \centering
   \includegraphics[width=14.5cm, angle=0]{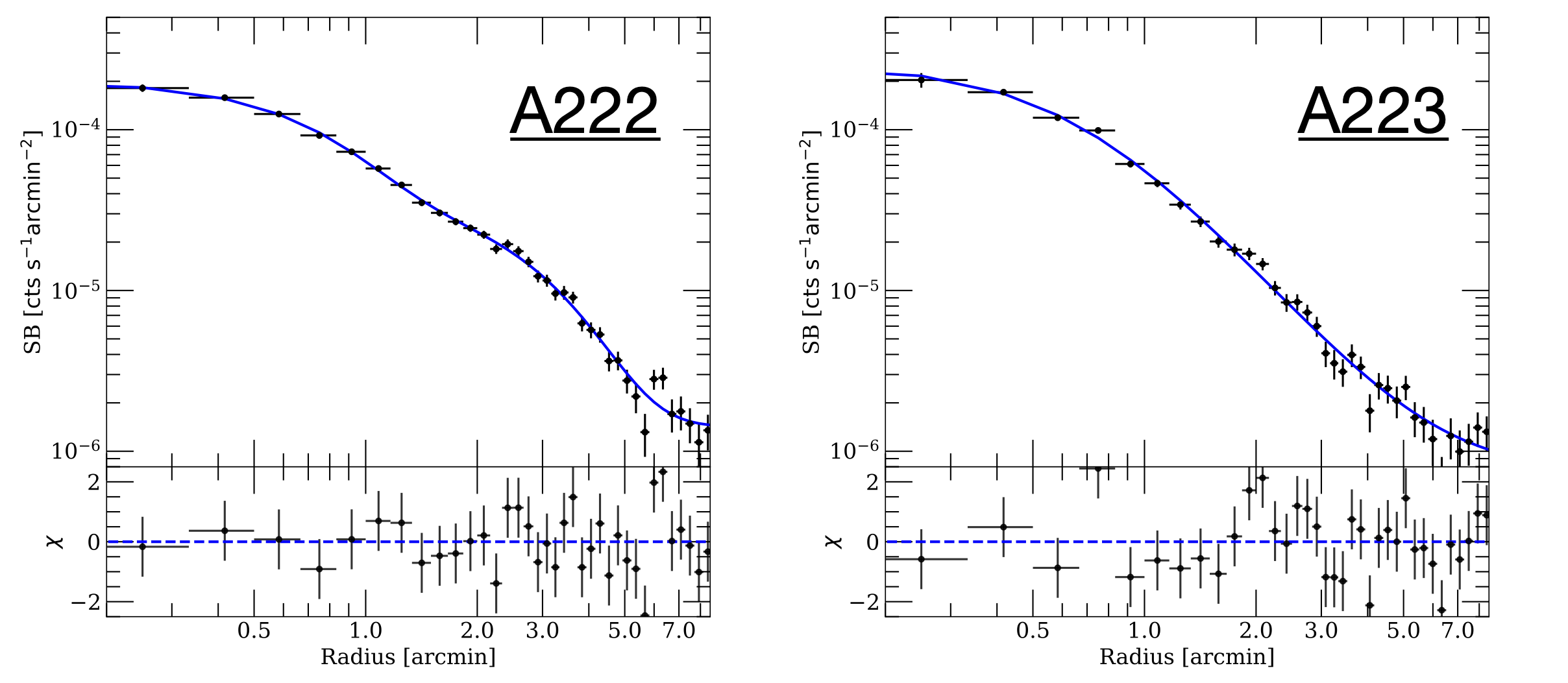}
   \caption{The surface brightness profile of A222 (left) and A223 (right). 
   The data were logarithmically rebinned, with the non-X-ray background subtracted. The best-fit double-beta models (see the main text) are shown in solid lines. The residuals are shown in the bottom panels. 
   }
   \label{fig:sbp-prof}
   \end{figure}

We used Chandra Interactive Analysis of Observations (CIAO) v4.15 and CALDB 4.11.2\footnote{https://cxc.cfa.harvard.edu/ciao/} to process the \textsl{Chandra} data. The event and auxiliary files were generated using the \texttt{chandra-repro} task, and all observations were merged. Stowed event files were used as particle background. The image is shown in Fig.\ref{fig:clus_scat}. To derive cluster surface brightness profiles, we defined regions (in cyan) that avoid the filament region and the A223 sub-cluster (see Appendix A).
The point sources detected (with \texttt{wavdetect}) are shown in green ellipses and are excluded from subsequent analyses. The surface brightness profiles (in the 0.5-7 keV band) are shown in Fig.~\ref{fig:sbp-prof} for A222 and A223, respectively.

\begin{figure} 
   \centering
   \includegraphics[width=15cm, angle=0]{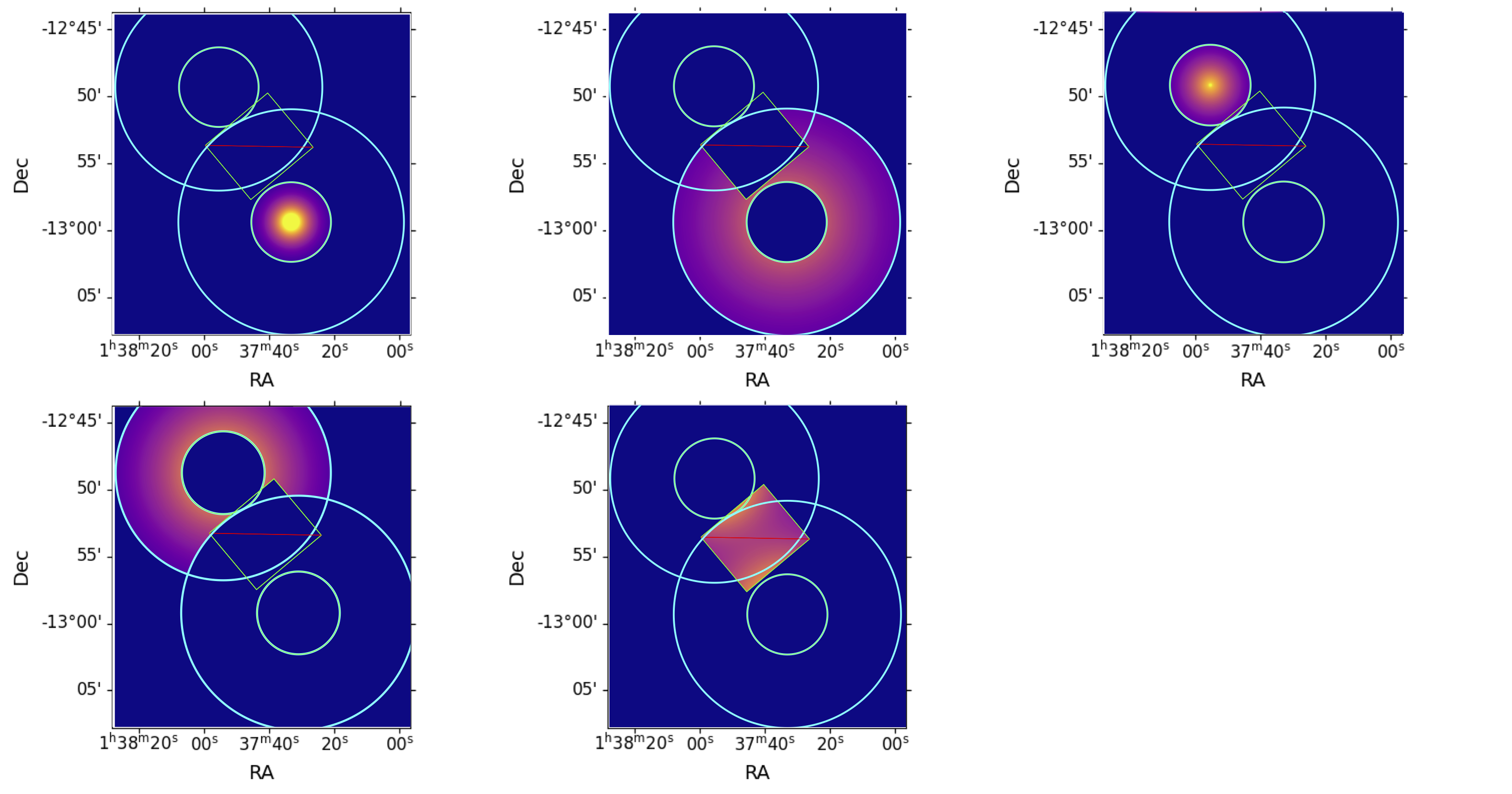}
   \put(-340,210){\makebox(0,0)[l]{{\color{white}\underline{a222-1}}}}
    \put(-200,210){\makebox(0,0)[l]{{\color{white}\underline{a222-2}}}}
    \put(-60,210){\makebox(0,0)[l]{{\color{white}\underline{a223-1}}}}
    \put(-340,100){\makebox(0,0)[l]{{\color{white}\underline{a223-2}}}}
    \put(-205,100){\makebox(0,0)[l]{{\color{white}\underline{filament}}}}
   \caption{Input images to the \textsl{XISSIM} simulation for the regions shown in the rows of Tab.~\ref{tab:clus_scat}. 
   The surface brightness distributions were derived from the \textsl{Chandra} surface brightness profiles. 
   }
   \label{fig:clus_scat_input}
   \end{figure}

The surface brightness profiles of the clusters were fitted with a double beta model by pyproffit \citep{Eckert20}: 
\begin{equation}
\label{eqn:beta}
    I\left(r\right) = I_{0}\left[\left(1+\left(x/r_{c,1}\right)^2\right)^{-3\beta+0.5} + R\left(1+\left(x/r_{c,2}\right)^2\right)^{-3\beta+0.5}\right] + B,
\end{equation}
where \(x\) is radial distance, and \(B\) represents the sky background. The best fits are shown in Fig.\ref{fig:sbp-prof}, with the $\chi^2$ values being 54 (39 dof) and 45 (39 dof) for A222 and A223, respectively, and the model parameters summarized in Tab.~\ref{tab:betamdl}. The best-fit surface brightness profiles were used to generate input images for the \texttt{XISSIM} simulator, and the results are shown in Fig.~\ref{fig:clus_scat_input}.  For the simulation, we adopted an absorbed APEC spectral model, with a temperature of 3\,keV, which is typical of galaxy clusters of \( M_{\rm tot} \sim 3 \times 10^{14} \, M_{\odot} \) (\citealt{finoguenov01, dietrich12}). We ran the simulation to collect \( 10^6 \) photons for XIS0, XIS1, and XIS3, respectively.

\begin{table}
\bc
\begin{minipage}[]{100mm}
\caption{The best-fit parameters of the double beta model\label{tab:betamdl}}\end{minipage}
\setlength{\tabcolsep}{1pt}
\small
 \begin{tabular}{lcccccc}
  \hline\noalign{\smallskip}
Clusters & $\beta$ & $r_{c,1}$ & $r_{c,2} $ & $R$ & log$_{10} I_0$ & log$_{10} B$\\
  \hline\noalign{\smallskip}
A222 & 0.47 $\pm$ 0.02 & 0.00 $\pm$ 0.05 & 0.56 $\pm$ 0.04 & 1.07 $\pm$ 0.15 & -3.67 $\pm$ 0.06 & -5.78 $\pm$ 0.12 \\
A223 & 0.59 $\pm$ 0.02 & 0.13 $\pm$ 0.04 & 0.65 $\pm$ 0.06 & 0.91 $\pm$ 0.20 & -3.54 $\pm$ 0.08 & -6.56 $\pm$ 0.30 \\
  \noalign{\smallskip}\hline
\end{tabular}
\tablecomments{0.86\textwidth}{$r_{c,1}$ and $r_{c,2} $ are in units of arcmin. $I_0$ and $B$ are in units of $\rm cts\ s^{-1} arcmin^{-2}$. 
}
\ec
\end{table}

\begin{table}
\bc
\begin{minipage}[]{100mm}
\caption{Scattered flux fraction (\%) 
\label{tab:clus_scat}}\end{minipage}
\setlength{\tabcolsep}{1pt}
\small
 \begin{tabular}{cccccc}
  \hline\noalign{\smallskip}
 Source region → & a222-1 & a222-2 & a223-1 & a223-2 & filament \\
Receiving region ↓  & & & & &\\
  \hline\noalign{\smallskip}
a222-1 & 84.61 & 13.21 & 0.10 & 0.08 & 2.77 \\
a222-2 & 12.50 & 80.73 & 0.70 & 3.47 & 12.41 \\
a223-1 & 0.10 & 0.10 & 90.64 & 15.29 &7.02 \\
a223-2 & 0.13 & 0.76 & 4.62 & 75.05 & 3.55\\
filament & 2.48 & 2.97 & 3.72 & 3.59 & 73.93\\
bkg & 0.18 & 2.23 & 0.2 & 2.52 & 0.32\\
  \noalign{\smallskip}\hline
\end{tabular}
\ec
\tablecomments{0.86\textwidth}{Shown is the fraction of the flux of a source region (as in Fig.~\ref{fig:clus_scat_input}) that is scattered to a ``receiving'' region or that of the remaining flux in the source region. Also see Fig.~\ref{fig:clus_scat_output}. }
\end{table}

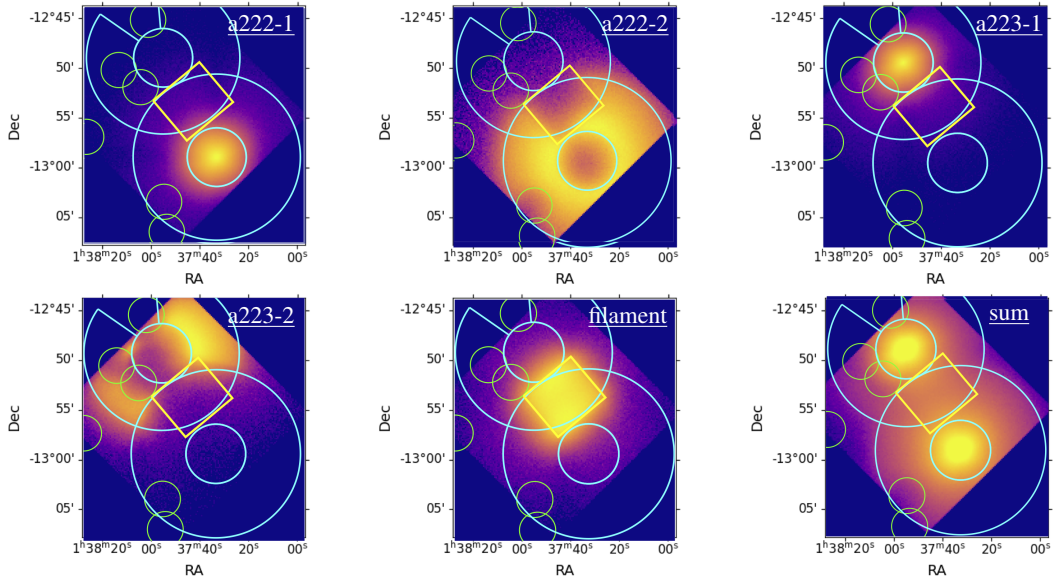
\begin{figure} 
   \centering
   \includegraphics[width=15cm, angle=0]{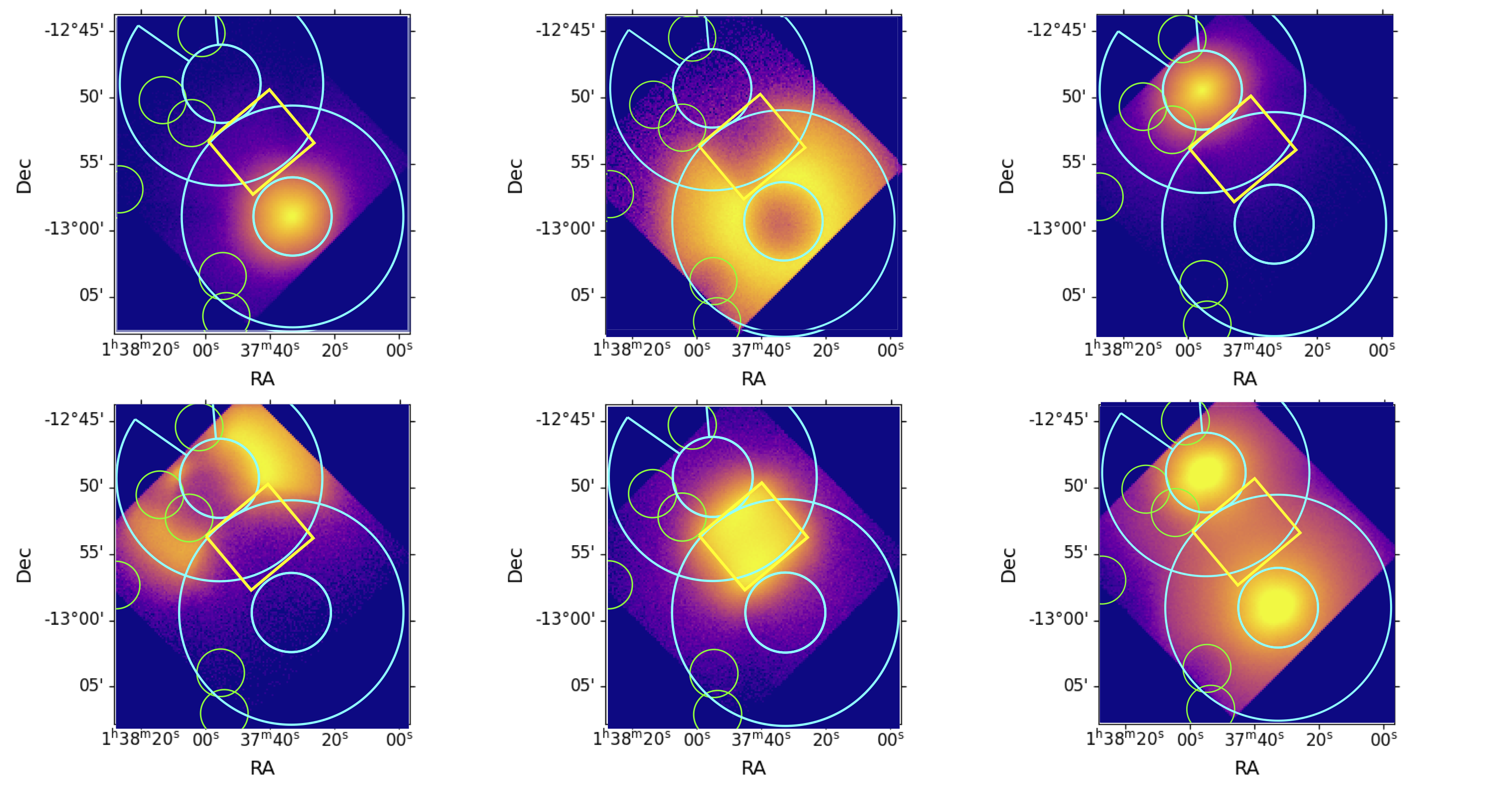}
  \put(-340,210){\makebox(0,0)[l]{{\color{white}\underline{a222-1}}}}
    \put(-200,210){\makebox(0,0)[l]{{\color{white}\underline{a222-2}}}}
    \put(-60,210){\makebox(0,0)[l]{{\color{white}\underline{a223-1}}}}
    \put(-340,100){\makebox(0,0)[l]{{\color{white}\underline{a223-2}}}}
    \put(-205,100){\makebox(0,0)[l]{{\color{white}\underline{filament}}}}
    \put(-55,100){\makebox(0,0)[l]{{\color{white}\underline{sum}}}}
   \caption{Simulated distribution of scattered emission. The distribution is shown for each of the regions of interest. The bottom-right panel shows the sum of all. } 
   \label{fig:clus_scat_output}
   \end{figure}

The output images from the simulation are shown in Fig.~\ref{fig:clus_scat_output}, which show the distribution of scattered X-rays from each of the regions of interest. 
Based on the images, we computed the fraction of the flux of a ``source'' region that is scattered to all the other ``receiving" regions,
and the results are summarized in Tab.~\ref{tab:clus_scat}. 
Clearly, the effects of scattered light are significant for all regions and must, therefore, be taken into account properly in subsequent analyses.

\section{Results}
\label{sect:results}

A spectrum was extracted from each of the regions of interest, as well as the sky-background region. 
It contains contributions from the non-X-ray background (NXB), the cosmic X-ray background (CXB), the Local Bubble (LHB), and possibly the Galactic halo (GH), which need to be carefully modeled, in order to reveal signals of scientific interest:
\begin{itemize}
\item \textit{Non-X-ray background: }The NXB spectrum was generated with the \texttt{xisnxbgen} task. Because the counts in the 10-14 keV band are dominated by NXB, we use them to scale the NXB spectrum. 
The uncertainties are estimated to around 3.6\% \citep{Tawa08}. 

\item \textit{Cosmic X-ray background: }The logN-logS function of AGN derived from the Chandra Deep Field South (CDF-S) survey was used to estimate the unresolved CXB contribution \citep{Lehmer12}. Here, we focused on the 2.0-8.0 keV band and computed the CXB surface brightness with the \texttt{cxbtools} tool \citep{cxbtool2017}. For spectral analyses, we adopted an absorbed power law to model CXB, with the photon index fixed at 1.41. We also computed uncertainties for the regions of interest, as shown in Tab.~\ref{tab:ups}, and considered three cases: CXB at the mean level ($1.26 \times 10^{-11}\ \rm ~ erg\ s^{-1}cm^{-2}deg^{-2}$, referred to as ``med" hereafter), 1$\sigma$ higher (``hi"), and 1$\sigma$ lower (``lo"). 

\begin{table}
\bc
\begin{minipage}[]{100mm}
\caption{The 1$\sigma$ uncertainties of unresolved CXB fluxes} \label{tab:ups}\end{minipage}
\setlength{\tabcolsep}{1pt}
\small
 \begin{tabular}{ccccccc}
  \hline\noalign{\smallskip}
 a222-1 & a222-2 & a223-1 & a223-2 & fila-box & bkg\\
  \hline\noalign{\smallskip}
 $2.68$ & $1.39$ & $2.75$& $2.17$& $2.39$ & $2.64$ \\
  \noalign{\smallskip}\hline
\end{tabular}
\ec
\tablecomments{0.86\textwidth}{in units of $10 ^{-12} ~\rm erg\ s^{-1}cm^{-2} deg^{-2} $, computed in the 2-8 keV band.}
\end{table}

\item \textit{The Local Bubble and Galactic Halo: }To constrain the LHB and possible GH contribution, we included the ROSAT All-Sky Survey (RASS) data \citep{Snowden97} in the fitting. 
We extracted an RASS spectrum from an annulus centered at the midpoint along the merger axis of the two clusters, with inner and outer radii of 1$^\circ$ and 2$^\circ$, respectively. We modeled the spectrum with an APEC (for LHB), an absorbed APEC (for GH) and a power law (for CXB), with chemical abundances fixed to \(1 Z_{\odot}\). While the preliminary temperature of LHB and GH were set to be 0.08\,keV and 0.2\,keV, the best-fit LHB temperature and GH temperature were found at $0.092\pm 0.012$\,keV and $0.196 \pm 0.007$\,keV , which in line with previous results (\citealt{Kuntz00,Yeung24} for LHB, and \citealt{Henley13,Yoshino09} for GH).

\item \textit{Scattered light: }
All regions of interest are affected by scattered light. For spectral analyses, we used an APEC to model such a component for each region, with the model parameters tied together for all regions. The normalizations of the model are properly scaled to follow the ratios shown in Tab.~\ref{tab:clus_scat}. For instance, the ratio of the scattered flux in a222-2 to the flux in a222-1 was determined by \( \frac{12.50/A_{2-2}}{84.61/A_{2-1}} \), where \( A_{2-1} \) and \( A_{2-2} \) represent the sky areas of a222-1 and a222-2, respectively.

\end{itemize}

The above components are all subject to interstellar absorption in the Milky Way, except for LHB. We adopted the TBabs model for the absorption, with $N_{\rm H_{tot}}$ fixed at $1.62 \times 10^{20}\ \rm cm^{-2}$, which is a weighted average based on the Swift survey\footnote{https://www.swift.ac.uk/analysis/nhtot/docs.php} (\citealt{willingale13}). The spectra extracted from the a223-2 and bkg regions show an absorption line feature roughly at 0.65\,keV. It appears in the XIS1 spectrum but not in the XIS0/3 spectra, it is likely an artefact caused by the Optical Blocking Filter (OBF) contamination in XIS1\footnote{http://heasarc.nasa.gov/docs/suzaku/analysis/abc/}. In subsequent spectral analyses, we simply excluded the feature (0.55-0.80 keV) from the XIS1 spectra. 

With the spectra optimally binned (\citealt{kaastra16}), we carried out joint spectral fits across all the regions (see Fig.\ref{fig:regions}), as well as different instruments, to disentangle the source and background/foreground components. The best-fit model parameters are shown in Tab.\ref{tab:fit_para} for the background/foreground components. The uncertainties were derived from MCMC analysis. 
 
\begin{table}
\bc
\begin{minipage}[]{100mm}
\caption{The parameters of the sky background model\label{tab:fit_para}}\end{minipage}
\setlength{\tabcolsep}{1pt}
\small
 \begin{tabular}{lcccccc}
  \hline\noalign{\smallskip}
Component & Model & Parameters & Values & Unit & Status  \\
  \hline\noalign{\smallskip}
Relative normalization & Const & RASS & $1/400\pi$ &  & Fixed & \\
                    &       & XIS0  &   1        &  & Fixed & \\
                    &       & XIS1/3&    $1.0\pm 0.1 $      &  & Free   & \\
Galactic absorption & TBabs & $\mathrm{N_H}$ & $1.62\times 10^{20}$ & $\rm cm^{-2}$ & Fixed & \\
LHB & APEC & temperature & $0.092 \pm 0.012$ &\,keV & Free \\
    &      & redshift &   0  &     & Fixed & \\
    &      & abundance &   1  & $Z_{\odot}$ & Fixed \\
    &      & norm & $5.39 \pm 0.16 \times10 ^{-4}$ &  & Free \\
GH  & TBabs $\times$ APEC & temperature & $0.191 \pm 0.007$  &\,keV & Free \\
    &      & redshift &   0  &     & Fixed \\
    &      & abundance &  1  & $Z_{\odot}$ & Fixed \\
    &      & norm & $1.46 \pm 0.06 \times 10^{-3}$ &  & Free \\
Unresolved CXB  & TBabs $\times$ Powerlaw & pho index & 1.41 &\,keV & Fixed\\
                &          & norm      & [lo/med/hi] &  & Fixed \\
  \noalign{\smallskip}\hline
 \end{tabular}
\ec
\tablecomments{0.86\textwidth}{norm is a dimensionless value defined as $\frac{10^{-14}}{4\pi\left[D_{A}\left(1+z\right)\right]^2\int n_e n_H dV}$ \footnote{https://heasarc.gsfc.nasa.gov/xanadu/xspec/manual/node134.html}, where $D_A$ is the angular diameter distance to the source (cm), $dV$ is the volume element ($\rm cm^3$), and $n_e$, $n_H$ are the electron and H densities ($\rm cm^{-3}$). The relative normalization reflects the difference in the spectral extraction area between RASS and Suzaku. $Z_\odot$ is the solar abundance using table \citet{asplund09}. }
\end{table}

\begin{figure} 
   \centering
   \includegraphics[width=14.0cm, angle=0]{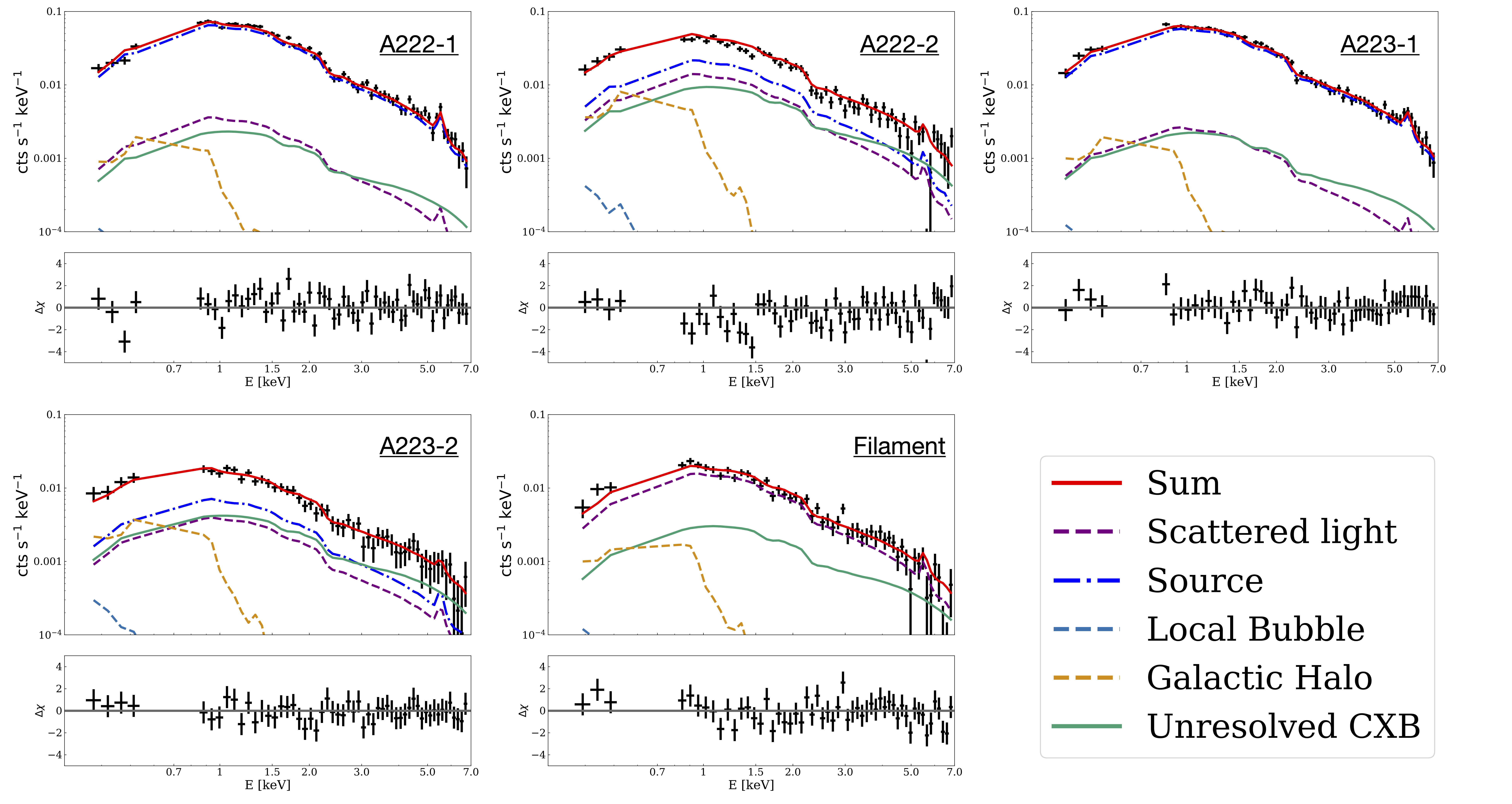}
   \caption{Observed X-ray spectra extracted from the regions of interest on XIS1 (see Fig.\ref{fig:regions}). Also shown in solid lines are the best-fit models, along with various model components.   } 
    \label{fig:spec-new}
    \end{figure}

Fig.\ref{fig:spec-new} displays the spectra and corresponding best-fit model for the regions of interest.
In the cluster outskirt regions (a222-2 and a223-2), CXB contributes most to the X-ray emission, followed by the intrinsic cluster emission and the scattered cluster light from the cores. The best-fit model parameters are shown in Tab.~\ref{tab:fit_clus-new}. The first set of uncertainties are the statistical errors derived from the MCMC analysis, while the second set are the systematic errors associated with the uncertainties in determining CXB contributions (which were derived by comparing fits in ``lo", ``mid", and ``hi" cases). 
The temperature is lower in the outskirt than in the core for A223, while it is comparable for A222, showing non cool-core characteristics, which is consistent with previous findings from \textsl{XMM-Newton} (\citealt{durret10}). The metallicity in the cores is found to be around 0.3$Z_{\odot}$, which is typical for galaxy clusters (\citealt{Walker19}), while it is poorly constrained in the outskirts (and thus fixed to 0.3$Z_{\odot}$, marked as 0.3$^*$ in the fitting process). 
For the filament region, the observed spectrum is well modeled by summing the background/foreground components, indicating that the data do not require a WHIM contribution.

\begin{table}
\bc
\begin{minipage}[]{100mm}
\caption{Best-fit gas properties in the cluster sectors\label{tab:fit_clus-new}}\end{minipage}
\setlength{\tabcolsep}{1pt}
\small
 \begin{tabular}{lcccc}
  \hline\noalign{\smallskip}
  & a222-1 & a222-2 & a223-1 & a223-2   \\
  \hline\noalign{\smallskip}
Temperature (keV) &$4.37_{-0.09}^{+0.10}~_{-0.08}^{+0.01}$ & $5.01_{-0.39}^{+0.45}~_{-0.12}^{+0.59}$ & $5.49_{-0.15}^{+0.18}~_{-0.26}^{+0.10}$& $5.27_{-0.70}^{+0.89}~_{-0.24}^{+0.07}$ \\
Abundance (solar) &$0.30_{-0.02}^{+0.02}~_{-0.02}^{+0.08}$ &
0.3 (fixed) & 
$0.42_{-0.07}^{+0.07}~_{-0.02}^{+0.03}$&
0.3 (fixed) \\
Normalization ($10^{-3}$)&
$95.73_{-0.91}^{+0.93}~_{-3.60}^{+3.96}$ & $8.01_{-0.22}^{+0.23}~_{-2.53}^{+3.08}$ & $91.12_{-0.90}^{+0.92}~_{-6.24}^{+1.97}$& $5.52_{-0.27}^{+0.26}~_{-3.44}^{+4.23}$ \\
C-statistics (dof) & 67(61) & 48 (54) & 68 (75) & 56 (68)  \\
  \noalign{\smallskip}\hline
 \end{tabular}
\ec
\tablecomments{0.86\textwidth}{Both statistical and systematic errors are shown (see the main text). The area-scaled normalization. 
}
\end{table} 

\begin{figure} 
   \centering
   \includegraphics[width=14.0cm, angle=0]{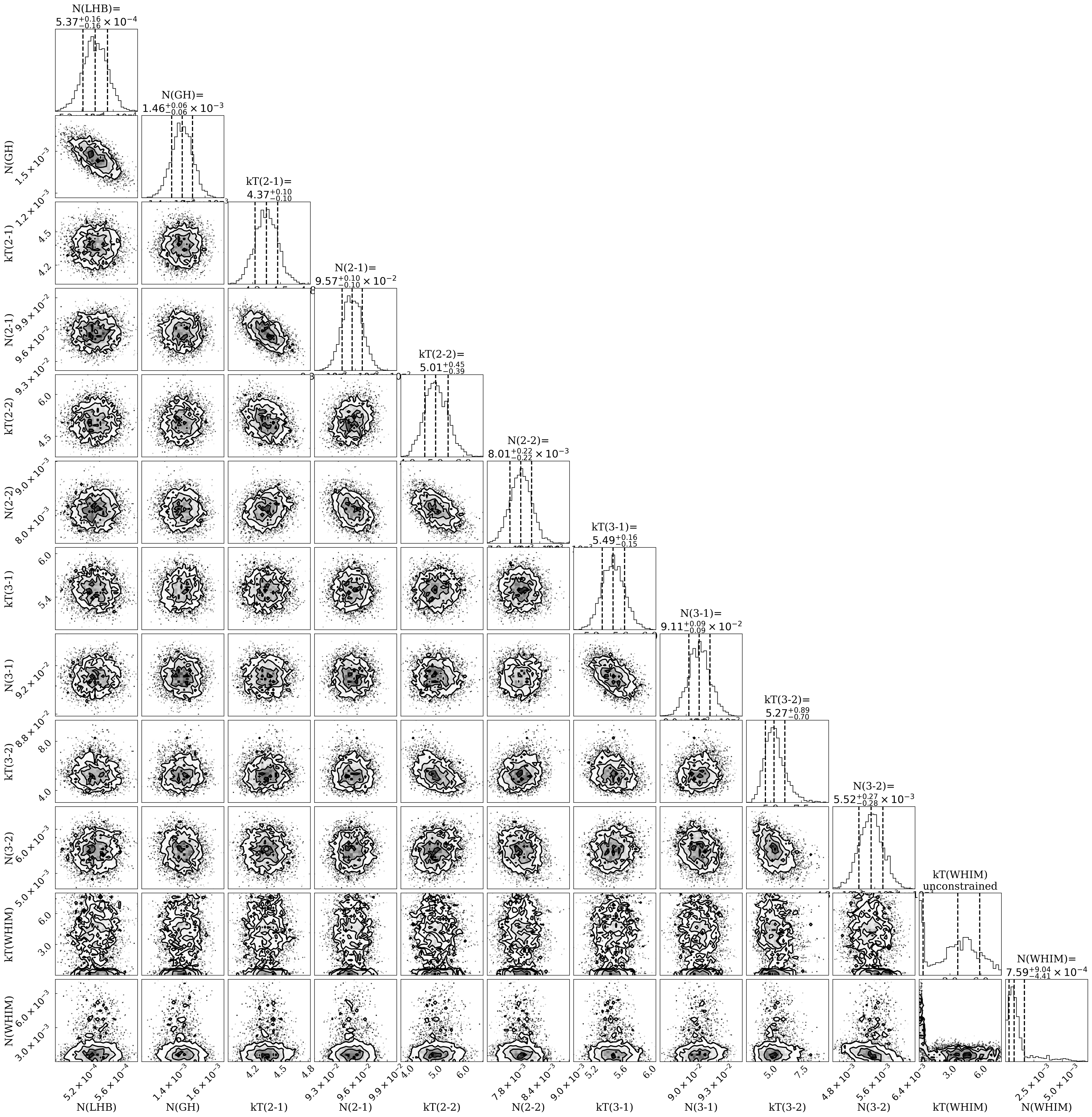}
   \caption{The parameter confidence intervals derived from MCMC analysis. APEC is adopted to model the emission from hot gas in all regions of interest. 
   In the name of a parameter, ``N" stands for normalization and ``kT" stands for temperature, with the numbers indicating the region of interest. For example, kT(2-1) shows the temperature of hot gas in region a222-1. The line plots on the top show the distributions of the parameters, with 1$\sigma$ confidence intervals indicated. kT is in units of keV, and normalization is dimensionless, as defined in Tab.\ref{tab:fit_para}. } 
    \label{fig:chain_new_all}
    \end{figure}

To derive upper limits on the properties of (undetected) WHIM emission, we refitted the spectra, with a new APEC component added to the model, while fixing the metallicities, the temperatures of GH and LHB, and the instrumental constants. We then carried out an MCMC analysis to determine confidence intervals for the remaining parameters, and the results are shown in Fig.\ref{fig:chain_new_all}. The temperature of the WHIM component is poorly constrained. The best-fit normalization is 7.59$\times 10^{-4}$, although it is less than 2$\sigma$ from 0. If we fixed the normalization at $3.5 \times 10^{-3}$ (as in \citealt{werner08}), we derived an upper limit on temperature to be 2.34\,keV (90\% confidence). On the other hand, if we fixed temperature at 0.91\,keV (as in \citealt{werner08}), we found the upper limit on the normalization to be $1.29\times10^{-3}$, corresponding to a density of $1.32 \times 10^{-5} \rm\ cm^{-3}$ ($\rho/\rho_c \sim 60$) (we followed \citealt{werner08} in converting normalization to density). 

\section{Discussion}
\label{sect:discussion}

Motivated by the detection of X-ray emission from a filament that connects Abell 222 and Abell 223 \citep{werner08}, we conducted a detailed study of the system with data from the \textsl{Suzaku} observations. The initial intention was to take advantage of the low NXB of the \textsl{Suzaku} detectors for detecting the extended emission of low surface brightness. However, we quickly ran into scattered-light issues, which add comparable uncertainties in determining surface brightness as the CXB. The results show that the emission detected in the said WHIM region  \citep{werner08} can be accounted for (within uncertainties) by the combination of the scattered cluster emission, CXB, Galactic foreground, and NXB. A WHIM component can be accommodated by the data, but its properties are not well constrained. With one parameter (normalization or temperature) fixed at the value in \citet{werner08}, the 90\% confidence upper limit on the other parameter is consistent with the published results. 

A possible way to reduce the CXB flux uncertainties in a \textsl{Suzaku} observation is to lower the source detection limit by also using data from observations with the telescopes of better angular resolution (e.g., \citealt{esra16, Zhang20}). To this end, we attempted to improve the constraints on the WHIM emission with the \textsl{XMM-Newton} data. The details of the data reduction are shown in Appendix A. In the end, we reduced the CXB flux and its uncertainty by approximately 30\%. We carried out similar spectral analyses to constrain the WHIM emission. The derived properties of cluster emission are consistent with those shown in Tab.~\ref{tab:fit_clus-discuss}, but those of the WHIM component are again poorly constrained. 
Fixing the WHIM temperature at kT=0.91\,keV, the upper limit on normalization is now $3.3\times 10^{-4}$, corresponding to $3.4 \times 10^{-6}\rm\ cm^{-3}$; fixing the normalization at $3.5 \times 10^{-3}$, the upper limit on kT is 0.16\,keV. Both of these upper limits are significantly lower than before. 

In summary, based on the \textsl{Suzaku} observation alone, we can neither confirm nor rule out the presence of the WHIM emission, but the derived upper limits on the WHIM properties are broadly in line with the published results \citep{werner08}. \citet{Mirahor22} also found evidence for X-ray emission from a filament that connects A2029 and A2033, using \textsl{Suzaku} data, and determined the WHIM temperature to be $1.4^{+0.7}_{-0.5}$\,keV, which is higher than our upper limit (derived with the help of \textsl{XMM-Newton} data). Even higher temperatures ($\gtrsim 3.5$\,keV) have been reported of WHIM emission in A399/401 and A3395/3391 \citep{Sakelliou04, Alvarez18, Reiprich21}, based on \textsl{eROSITA}, which were attributed to shock heating in those systems. 
The APEC normalization of the WHIM component (derived with both kT and normalization free) implies a 90\% upper limit of 101 on the WHIM over-density, which aligns with previous results. The WHIM overdensity was determined to be $\rho/\rho_c \lesssim 60$ for A2801/2804/2811, \citep{Sato10} and $\sim 160 \pm 70$ (showing 90\% confidence interval) for A2029/2033 \citep{Mirahor22}. 

\begin{acknowledgements}
The authors thank the anonymous referee for their comments that have improved this paper. The authors wish to thank Jelle de Plaa, Jelle Kaastra, Chen Li, Zhenlin Zhu, Anwesh Majumder, Xiaoyuan Zhang, Dimitris Chatzigiannakis, Thomas Plsek, Jean-Paul Breuer, Jiejia Liu, Yu Zhou, Hiroki Akamatsu, Thomas Reiprich, Nobert Werner, Dominique Eckert, Junjie Mao, and Joop Schaye for helpful discussion, and the Tsinghua Astrophysics High-Performance Computing platform for providing computational and data storage resources. This work was supported in part by the National Natural Science Foundation of China through Grants 11821303, and by the Ministry of Science and Technology of China through Grant 2018YFA0404502. YLC acknowledges support from the China Scholarship Council, and DH acknowledges the financial support of the GA\v{C}R EXPRO grant No. 21-13491X. This research is based on observations obtained
from the Suzaku satellite, a collaborative mission between the space agencies of
Japan (JAXA) and the USA (NASA). This paper employs a list of Chandra datasets, obtained by the Chandra X-ray Observatory, contained in the Chandra Data Collection (CDC) ~\href{https://doi.org/10.25574/cdc.371}{DOI: 10.25574/cdc.371}.
\end{acknowledgements}

\appendix

\section{Deriving unresolve CXB flux with \textsl{XMM-Newton} observations}

\begin{figure} 
   \centering
   \includegraphics[width=14.0cm, angle=0]{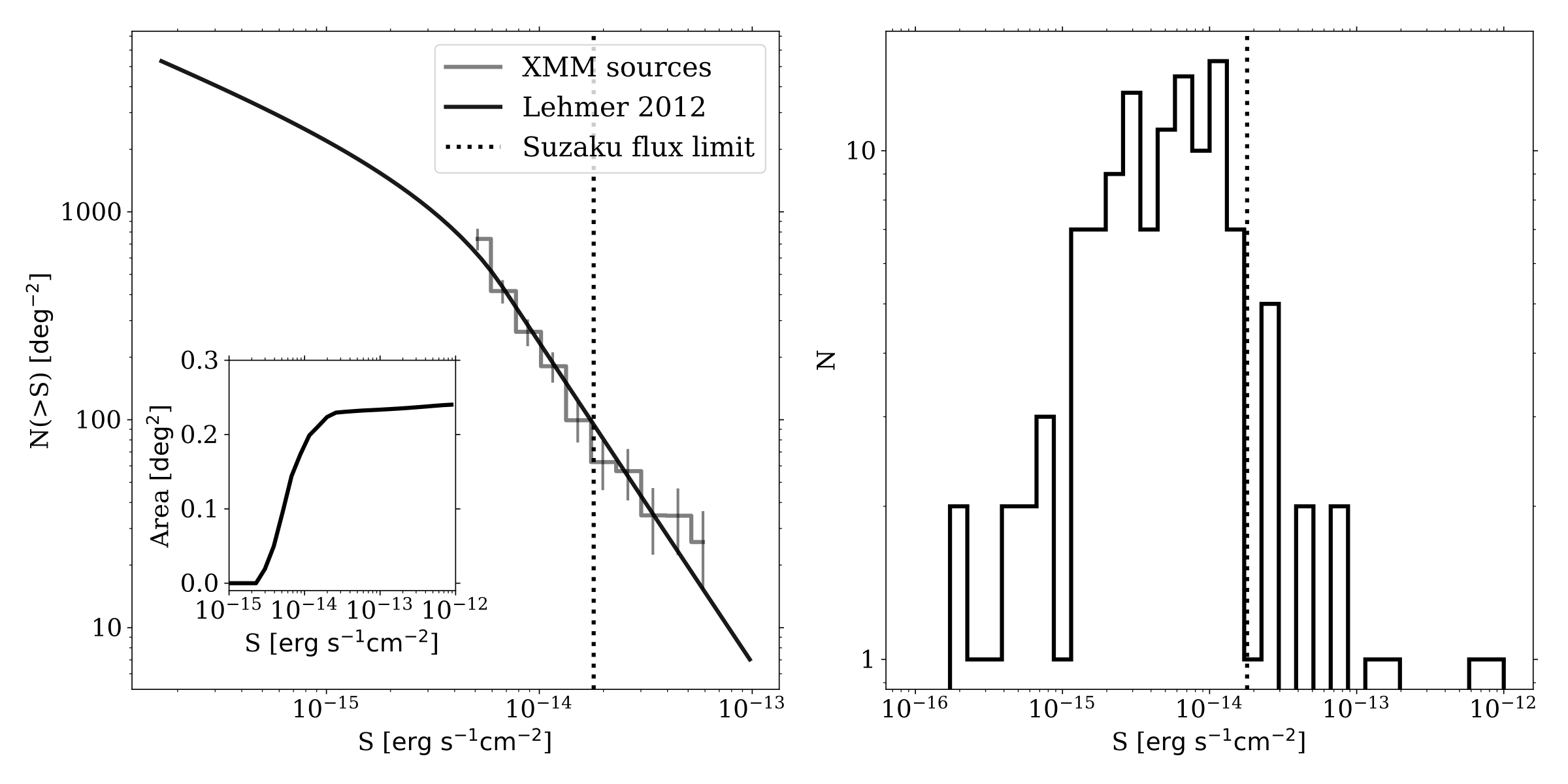}
   \caption{Left: The logN-logS curve of the \textsl{XMM-Newton}, the grey line is number of point sources whose fluxes higher than x-axis flux, corrected for the effective area. The black curve is the logN-logS curve from \citet{Lehmer12}. Right: The histogram of point sources. The dotted line is the \textsl{Suzaku} detection limit. }
   \label{fig:ups}
   \end{figure}

In previous study using \textsl{Suzaku} data, a uniform detection limit has been applied across the entire FOV (e.g. \citealt{Walker12, esra16}). However, this approach does not utilize all the resolved CXB information. To address this, we calculated the unresolved point source flux on a pixel-by-pixel basis using the \textsl{XMM-Newton} sensitivity map. We followed the method outlined in \citet{Huang24} to generate a sensitivity map in the 2.0–4.5 keV band to determine the detection limit of every pixel. The CDF-S logN-logS aligns with A222-223 FOV (see Fig.\ref{fig:ups}).  

\begin{table}
\bc
\begin{minipage}[]{100mm}
\caption{Resolved CXB flux and the $1 \sigma$ errors
\label{tab:rps-discuss}}\end{minipage}
\setlength{\tabcolsep}{1pt}
\small
 \begin{tabular}{lccccc}
  \hline\noalign{\smallskip}
a222-1 & a222-2 & a223-1 & a223-2 & fila-box & bkg\\
  \hline\noalign{\smallskip}
$0.84 \pm 0.14$ & $2.22 \pm 0.37$ & $3.25 \pm 0.54$ & $2.85 \pm 0.47$ & $4.41 \pm 0.73$ & $0.15 \pm 0.03 $ \\
  \noalign{\smallskip}\hline
\end{tabular}
\ec
\tablecomments{0.86\textwidth}{The photon index is 1.89. The flux is in the 2.0-8.0\,keV band. The flux unit is $10 ^{-12} \rm erg\ s^{-1}cm^{-2} deg^{-2} $. }
\end{table}

\begin{table}
\bc
\begin{minipage}[]{100mm}
\caption{Unresolved CXB fluxes and the $1 \sigma$ errors \label{tab:ups-discuss}}\end{minipage}
\setlength{\tabcolsep}{1pt}
\small
 \begin{tabular}{lccccc}
  \hline\noalign{\smallskip}
a222-1 & a222-2 & a223-1 & a223-2 & fila-box & bkg\\
  \hline\noalign{\smallskip}
$9.14 \pm 1.42$ & $7.68 \pm 0.63$ & $7.52 \pm 1.24$& $8.06 \pm 0.95$& $7.25 \pm 0.83$ & $8.37 \pm 1.24$ \\
  \noalign{\smallskip}\hline
\end{tabular}
\ec
\tablecomments{0.86\textwidth}{ The flux is in the 2.0-8.0 keV band. The flux unit is $10 ^{-12} \rm erg\ s^{-1}cm^{-2} deg^{-2} $}
\end{table}

\begin{figure} 
   \centering
   \includegraphics[width=14.0cm, angle=0]{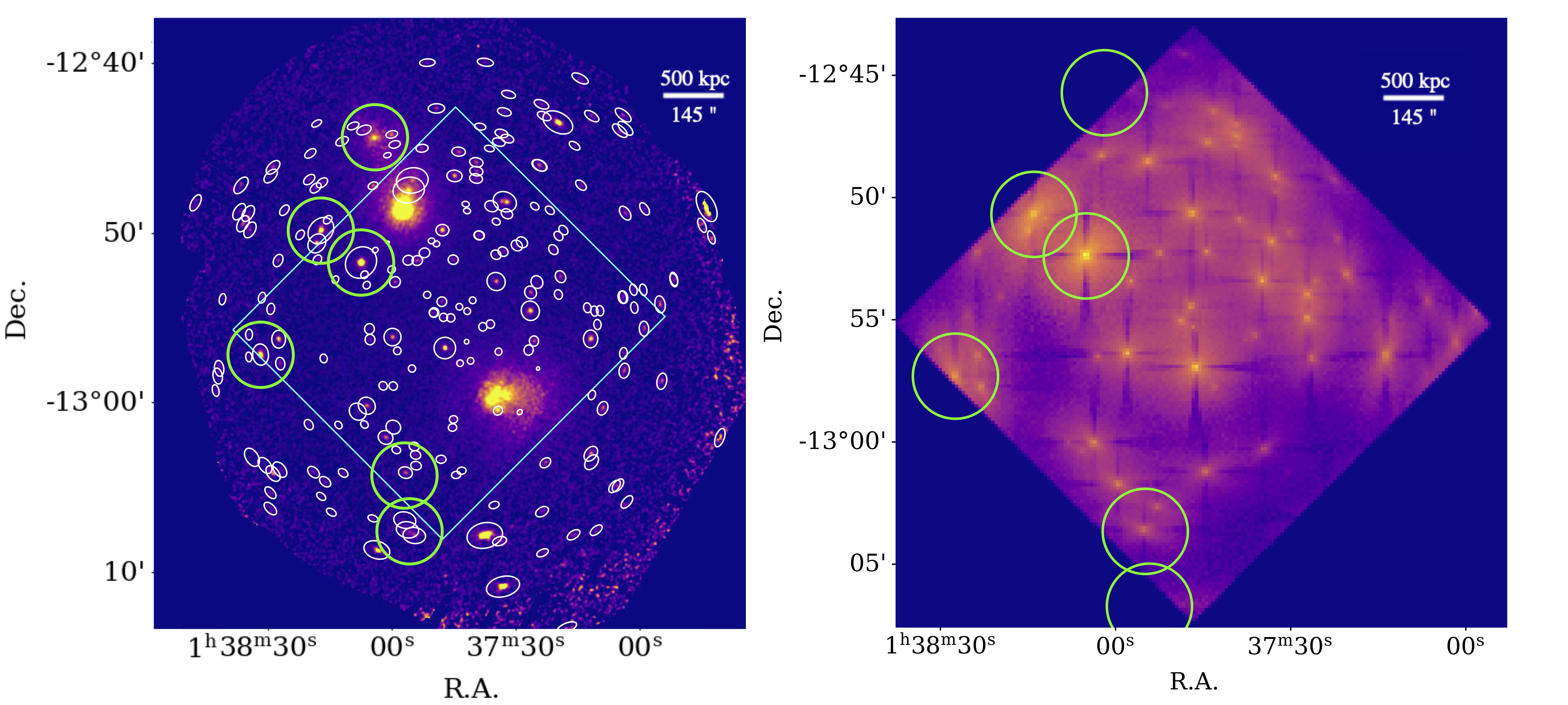}
   \caption{Left: The 0.5-7.0 keV mosaic image from \textsl{XMM-Newton}, smoothed by a Gaussian kernel of FWHM 10$''$. White ellipses are the point sources resolved by \textsl{XMM-Newton}. Green circles are the point source exclusion regions for \textsl{Suzaku}. Right: The output XISSIM map of \textsl{XMM-Newton} resolved point sources in XIS1 in the 0.5-12.0 keV band. } 
   \label{fig:bkg_sum}
   \end{figure}

To calculate the CXB resolved by \textsl{XMM-Newton}, we considered the scattering effects of point sources. First, we selected sources from the \textsl{XMM-Newton} catalogue and ran \texttt{XISSIM} simulations for each source, all with the same input photon number. Then we used a calibrated on-axis source to account for vignetting effect, and scale output image with source fluxes in \textsl{XMM-Newton} catalogue. Lastly, we summed all the output images, which was then used in subsequent analyses. In Fig.\ref{fig:bkg_sum}, the resolved point sources are marked in \textsl{XMM-Newton} image (white). In the \textsl{XMM-Newton} image, we can see the northeastern sub-cluster associated with A223 is visible in both \textsl{XMM-Newton}, but is undetected in the \textsl{Suzaku} observations due to the limited FOV. The centroid of this sub-cluster is located at a projected angular separation of 4.8$'$ from the A223 cluster center, corresponding to a physical separation of approximately 987\,kpc, which is positioned at (01:38:01.8, -12:45:07.0) \citep{dietrich02}. Separating the subcluster from the main cluster is difficult, so the subcluster mass was not measured in previous works \citep{dietrich02, dietrich05}. The mass density map from weak lensing analysis is available in \citealt{dietrich12}.

Using the updated CXB tables of resolved (Tab.\ref{tab:rps-discuss}) and unresolved CXB (Tab.\ref{tab:ups-discuss}), we refitted the spectra. The CXB resolved by \textsl{XMM-Newton} was additionally modelled by a power law. The photon index is constrained at $1.89 \pm 0.34$ by fitting \textsl{XMM-Newton} sky background spectra. We considered the photon index uncertainties, and the normalization is ensured to be the same. Tab.\ref{tab:fit_clus-discuss} shows the spectral fitting results. The properties are in line with the results in Section \ref{sect:results} in $1\sigma$.

\begin{table}
\bc
\begin{minipage}[]{100mm}
\caption{The gas properties of the cluster sectors
\label{tab:fit_clus-discuss}}\end{minipage}
\setlength{\tabcolsep}{1pt}
\small
 \begin{tabular}{lcccc}
  \hline\noalign{\smallskip}
  & a222-1 & a222-2 & a223-1 & a223-2   \\
  \hline\noalign{\smallskip}
Temperature (keV)&
$4.46_{-0.10}^{+0.11}~_{-0.06}^{+0.18}~_{-0.04}^{+0.21}$ & $5.89_{-0.56}^{+0.85}~_{-0.03}^{+0.97}~_{-1.03}^{+1.68}$ & $5.70_{-0.15}^{+0.19}~_{-0.18}^{+0.13}~_{-0.41}^{+0.15}$& $6.40_{-1.00}^{+1.26}~_{-0.52}^{+1.23}~_{-1.37}^{+1.90}$ \\
Abundance (solar)&$0.33_{-0.02}^{+0.02}~_{-0.05}^{+0.02}~_{-0.03}^{+0.02}$ &
0.3$^*$ & 
$0.42_{-0.07}^{+0.07}~_{-0.01}^{+0.01}~_{-0.03}^{+0.13}$&
0.3$^*$ \\
Normalization ($10^{-3} $)&$97.31_{-1.10}^{+1.00}~_{-2.99}^{+7.45}~_{-7.32}^{+14.06}$ & $6.84_{-0.22}^{+0.20}~_{-0.88}^{+1.41}~_{-2.18}^{+3.33}$ & $88.23_{-0.80}^{+0.81}~_{-2.21}^{+3.56}~_{-9.48}^{+10.64}$& $5.11_{-0.21}^{+0.22}~_{-2.81}^{+1.40}~_{-4.88}^{+10.23}$ \\
C-statistics (dof) & 65(60) & 50(48) & 45(60) & 36(55) \\
  \noalign{\smallskip}\hline
 \end{tabular}
\ec
\tablecomments{0.86\textwidth}{The first errors presented in the results are the statistical errors, the second errors represent the systematic errors of different unresolved CXB levels, and the third errors refer to the systematic errors of different photon index of resolved CXB. The C-statistics here refer to the fitting statistics of XIS1 spectra.} 
\end{table}

\bibliographystyle{raa}
\bibliography{bibtex}
\label{lastpage}

\end{document}